# On the Metric of Space-Time

**Carl E. Wulfman**[*]
Department of Physics, University of the Pacific, Stockton, USA
Email: wulfmanc@wavecable.com

## ABSTRACT

Maxwell's equations are obeyed in a one-parameter group of isotropic gravity-free flat space-times whose metric depends upon the value of the group parameter. An experimental determination of this value has been proposed. If it is zero, the metric is Minkowski's. If it is non-zero, the metric is not Poincare invariant and local frequencies of electromagnetic waves change as they propagate. If the group parameter is positive, velocity-independent red shifts develop and the group parameter plays a role similar to that of Hubble's constant in determining the relation of these red-shifts to propagation distance. In the resulting space-times, the velocity-dependence of red shifts is a function of propagation distance. If 2c times the group parameter and Hubble's constant have approximately the same value, observed frequency shifts in radiation received from stellar sources can imply source velocities quite different from those implied in Minkowski space. Electromagnetic waves received from bodies in galactic Kepler orbits undergo frequency shifts which are indistinguishable from shifts currently attributed to dark matter and dark energy in Minkowski space, or to a non-Newtonian physics.

[*] Professor emeritus

## 1. Introduction

Because cosmology so heavily depends upon interpretations of the properties of electromagnetic radiation, any misunderstandings of these properties can have extensive consequences. Maxwell's equations govern the propagation and modification of this radiation and permit relations between observations of cosmic events to be codified in global metrics which determine geometric relations in space-time.

These geometric relations provide the framework for the analyses in this paper, which is based on work of Harry Bateman, Ebenezer Cunningham, and E.L. Hill. We show that Minkowski's metric cannot be assumed, *a priori*, to be the metric of physical gravitation-free space-time. It is just one member of a one-parameter family of allowed metrics. As a consequence, current observations allow Hubble's relations to have consequential new interpretations. Among these are interpretations that eliminate the current need for dark matter and dark energy.

In 1909, Bateman [1] and Cunningham [2] proved that Maxwell's equations are invariant under an inversion, and the transformations of a fifteen-parameter Lie group. Thirty-six years later, just as WW II ended, E. L. Hill showed that the infinitesimal transformations of a one-parameter subgroup of the Bateman-Cunningham Lie group define a relation isomorphic to Hubble's law [3].

Hill's discovery was overlooked for 65 years. In the year 2000, unaware of it, Hoyle, Burbidge, and Narlikar used the Bateman-Cunningham Lie group in their development of a generalized relativistic cosmology [4], and in 2010, Tomilchik [5] argued that Hubble's relations are a consequence of properties of transformations of the special conformal subgroup of the Bateman-Cunningham group. Hill's paper is however listed in Kastrup's recent review of papers dealing with conformal invariance [6], which led to the research reported here, and to the realization that the value of Hill's group parameter can be determined by experiments proposed by Wulfman [7,8]. A more recent list of references to research on the physical role of conformal transformations is contained in an article on the quantum mechanics of accelerated relativistic particles written by Calixto, *et al*. [9].

Section 2 of the present paper describes the finite transformations that result from the infinitesimal transformations investigated by Hill. In determining these transformations and their effects on electromagnetic waves and Minkowski's metric, we make use of properties of Lie transformation groups that can be found in recent texts by Cantwell [10], by Hydon [11], by Stephani [12], and by Wulfman [13].

In Section 3, it is shown that the transformations convert the usual plane-wave and spherical-wave solutions of Maxwell's equations into one-parameter families of solutions with local wave-numbers that evolve as the waves propagate.

The action of the transformations on Minkowski's metric is determined in Section 4. They convert Minkowski's metric into a one parameter family of metrics which define a one parameter family of space-times. It is shown that for each value of the group parameter the transformed waves propagate in a transformed space-time. whose metric is fixed by the value of the group parameter. The transformed Minkowski metrics are not Poincare invariant.

In Section 5, to develop physical consequences of the mathematical connection between the metric of the space-times and the behavior of electromagnetic radiation in them, local measurements are expressed in terms of coordinates of points in Minkowski space, and in terms of the transformed coordinates produced by the action of the group. Locally observed frequency shifts in the one-parameter family of space-times are thereby related to "Doppler shifts" in Minkowski space. The development of velocity-independent frequency shifts in the transformed space-times leads to fundamental revisions of current interpretations of Hubble's relations.

Then, in Section 6, it is shown that the transformed space-times have a surprising property: in them, electromagnetic waves received from bodies in galactic Kepler orbits can undergo frequency shifts which are indistinguishable from the shifts currently attributed to the hypothesized presence of dark matter and dark energy in Minkowski space.

The final section of the present paper contains a brief discussion dealing with the relation between the metrics considered here, and those applicable in the presence of gravitational fields.

## 2. The Isotropic Transformations of the Special Conformal Group

Hill's one-parameter group of isotropic special conformal transformations inter-converts the vectors **X**, **X'**, and *r*, *r'* with

$$\mathbf{X} = (x^1, x^2, x^3, x^4), x^4 = ct \quad (1,2)$$

$$\mathbf{X'} = (x'^1, x'^2, x'^3, x'^4), x'^4 = ct', \quad (3,4)$$

and

$$\mathbf{r} = (x^1, x^2, x^3) \quad (5)$$

$$\mathbf{r'} = (x'^1, x'^2, x'^3) \quad (6)$$

The italic superscripts denote locally Cartesian components. Here, and in some of the following paragraphs, italics are also used to call attention to variables such as $r$ and $x^4$ that may take on initial values, final values, and all values in between, as the group parameter varies from zero to its final value, except we will use "v" to refer to velocity and "ν" to refer to frequency. The Lie generator of the group can be expressed as

$$C_4 = \left(r^2 + (x^4)^2\right) \partial/\partial x^4 + 2x^4 r \partial/\partial r. \quad (7)$$

The transformations can be carried out by the operator $\exp(\gamma C_4)$. It produces the projective transformations

$$(x^4 - r) \to x'^4 - r' = \exp(\gamma C_4)(x^4 - r) \quad (8)$$

$$= (x^4 - r)/(1 - \gamma(x^4 - r)) \quad (9)$$

and

$$(x^4 + r) \to x'^4 + r' = \exp(\gamma C_4)(x^4 + r) \quad (10)$$

$$= (x^4 + r)/(1 - \gamma(x^4 + r)). \quad (11)$$

Acting on $r, x^4$, $\exp(\gamma C_4)$ produces the transformations

$$r \to r' = r/D, \quad (12)$$

$$x^4 \to x'^4 = \left(x^4 - \gamma(x^4 - r^2)\right)/D \quad (13)$$

with

$$D = (1 - \gamma x^4)^2 - (\gamma r)^2 \quad (14)$$

$$= (1 - \gamma(x^4 + r))(1 - \gamma(x^4 - r)) \quad (15)$$

The inverse transformations are

$$(x'^4 - r') \to x^4 - r = \exp(\gamma C_4)(x^4 - r) \quad (16)$$

$$= (x'^4 - r')/(1 + \gamma(x'^4 - r')) \quad (17)$$

and

$$(x'^4 + r') \to x^4 + r = \exp(\gamma C_4)(x^4 + r) \quad (18)$$

$$= (x'^4 + r')/(1 + \gamma(x'^4 + r')). \quad (19)$$

and

$$r' \to r = r'/D', \quad (20)$$

$$x'^4 \to x^4 = \left(x'^4 + \gamma((x'^4)^2 - r'^2)\right)/D', \quad (21)$$

$$D' = (1 + \gamma x'^4)^2 - (\gamma r')^2 \quad (22)$$

$$= (1 + \gamma(x'^4 + r'))(1 + \gamma(x'^4 - r')). \quad (23)$$

Changing $\gamma$ to $-\gamma$ and also interchanging $r$, $r'$ and $x^4$, $x'^4$ produces the inverse transformation.

The transformations of the Bateman-Cunningham group are conformal in the sense that they do not alter angles between spatial vectors. The transformations generated by $C_4$ also do not alter the origin of coordinates. They interconvert

$$\mathbf{X} = (r\sin(\theta)\cos(\Phi), r\sin(\theta)\sin(\Phi), r\cos(\theta), x^4) \quad (24)$$

and

$$\mathbf{X}' = (r'\sin(\theta)\cos(\Phi), r'\sin(\theta)\cos(\Phi), r'\cos(\theta), x'^4). \quad (25)$$

If $X$ has its origin at the vertex of a light cone where a source is located, then $\mathbf{X}'$ will originate at the same source. If the time coordinates have non-negative values, we will say the coordinates are those of points on "source cones" on which $x^4 = r$ and $x'^4 = r'$. See Figure 1. A continuously emitting source produces a continuum of light cones with a common time axis. The vertex of any one of these may be considered that of a source cone.

FIGURE 1

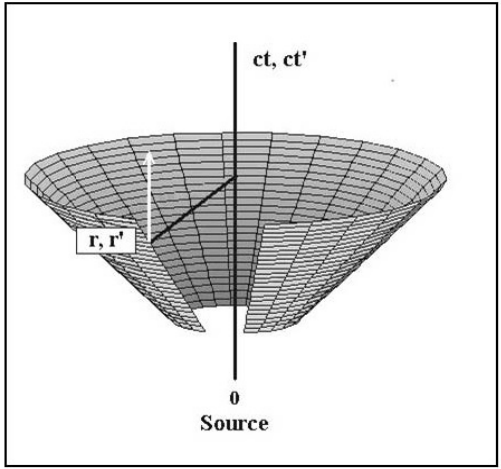

Light cone showing the trajectory of an observer (white vector), whose local velocity with respect to the light source is zero.

The first extension of $C_4$ is

$$C_4^{[1]} = C_4 + 2r(1 - v_r^2)\partial/\partial v_r, \quad (26)$$

$$v_r = dr/dx^4. \quad (27)$$

It generates transformations of radial velocities in which:

$$v_r \to v'_r = (Av_r + B)/(A + Bv_r). \quad (28)$$

The inverse of this is

$$v'_{r'} \to v_r = (Av'_{r'} - B)/(A - Bv'_{r'}).. \quad (29)$$

In these relations

$$A = \partial r'/\partial r = \partial x'^4/\partial x^4 = (1 + \gamma x'^4)^2 + (\gamma r')^2 \quad (30\text{-}32)$$

$$= ((1 - \gamma x^4)^2 + (\gamma r)^2)/D^2, \quad (33)$$

$$B = \partial r'/\partial x^4 = \partial x'^4/\partial r = 2(\gamma r')(1 + \gamma x'^4) \quad (34\text{-}36)$$

$$= 2(\gamma r)(1 - \gamma x^4)/D^2. \quad (37)$$

D is the function defined in (15). Because of the $(1 - v_r^2)$ factor in $C_4^{[1]}$, the effects of these transformations weaken as $v_r$ approaches the speed of light, 1, which they cannot alter. Because both $C_4$ and $C_4^{[1]}$ are quadratic functions of $\gamma r$ and $\gamma x^4$, they generate transformations that are a weak function of both the coordinates and the group parameter near sources, terrestrial as well as celestial.

As $C_4$ annihilates,

$$I_1 = ((x^4)^2 - r^2)/r \quad (38)$$

the transformations generated by it leave this function invariant. The transformations generated by $C_4^{[1]}$ also leave invariant the velocity-dependent function

$$I_2 = \{(x^4 - r)/(x^4 + r)\}((1 + v_r)/(1 - v_r))^{1/2}. \quad (39)$$

This invariant has implications which will be used in Section 7.

Because $C_4$ does not commute with any of the generators $\partial/\partial x^k$, the transformations generated by it and $C_4^{[1]}$ are not translation invariant. A translation that moves the origins of the vectors $\mathbf{X}$ from $\mathbf{O}$ to

$$A = (a^1, a^2, a^3, a^4) \quad (40)$$

converts $\mathbf{X}$ to $\mathbf{X} + \mathbf{A}$. It moves the origin of the vectors $X'$ from $\mathbf{O}$ to

$$\mathbf{X}'_A = \mathbf{X}' + \mathbf{A}, \quad (41)$$

which is not the same as $(\mathbf{X} + \mathbf{A})'$.

## 3. Electromagnetic Waves with Evolving Wave Numbers

In this section, we first deal with plane waves, the waves that are potentially observable at great cartesian distances from sources. An electromagnetic plane wave with wave number k propagating in the direction of $x^3$ may be defined by the equation

$$\boldsymbol{\Psi} = (\mathbf{E} + \mathbf{H})\cos(k(x^3 - x^4) + \Phi_0), \quad (42)$$

with

$$\mathbf{E} \cdot \mathbf{x}^3 = 0, \mathbf{H} \cdot \mathbf{x}^3 = 0, \quad (43,44)$$

$$\mathbf{E} \cdot \mathbf{H} = 0, |\mathbf{E}| = |\mathbf{H}| = \text{const}. \quad (45\text{-}47)$$

Both curl (**E**) and curl (**H**) vanish, and the function $\cos(k(x^3-x^4)+\Phi_0)$ satisfies the wave equation, which reduces to the requirement that

$$(\partial/\partial x^3 - \partial/\partial x^4)(\partial/\partial x^3 + \partial/\partial x^4)$$

annihilate  (48)

$$\cos(k(x^3-x^4)+\Phi_0)$$

Spherical waves with the same wave number are obtained if the cosine function is replaced by $\sin(k(r-x^4))/(r-x^4)$.

The inverse of (9) produces the relation

$$x^3 - x^4 = (x'^3 - x'^4)/(1+\gamma(x'^3 - x'^4)) \qquad (49)$$

Consequently, in the **X**' system, the wave defined by (42) becomes defined by

$$\Psi'(x'^3, x'^4) = (\mathbf{E}' + \mathbf{H}')\cos\left(k\left(\frac{(x'^3 - x'^4)}{1+\gamma(x'^3 - x'^4)}\right) + \Phi_0\right)$$

(50)

with

$$|\mathbf{H}'| = |\mathbf{E}'|, \mathbf{E}' \neq \mathbf{E}. \qquad (51,52)$$

One thus need only require that $\Psi'$ continues to satisfy the wave equation. Equation (48) ensures that it does so everywhere that the denominator does not vanish. Furthermore, because $\text{curl}(\mathbf{E}')$ and $\text{curl}(\mathbf{H}')$ vanish, no external potentials or currents are required to produce the EM waves defined by (42) and those defined by (50).

The wave that carries the fields can be considered to have wave number

$$k' = k/(1+\gamma(x'^3 - x'^4)). \qquad (53)$$

At a stationary point with coordinate $x_0'^3$ the waves oscillate with a frequency

$$\omega = k'c = kc/(1+\gamma(x_0'^3 - x'^4)). \qquad (54)$$

The corresponding spherical waves are defined by

$$\Psi'(r', x'^4) = \left\{\frac{1+\gamma(r'-x'^4)}{(r'-x'^4)}\right\} \sin\left\{\frac{(k(r'-x'^4))}{1+\gamma(r'-x'^4)}\right\} \qquad (55)$$

They carry electromagnetic fields in which **E**' and **H**' are everywhere orthogonal to the wave-front and to each other.

## 4. The One-Parameter Family of Metrics

Differentiation of the finite $C_4$ transformation (29) produces the relations

$$dr = Adr' + Bdx'^4, \qquad (56)$$

$$dx^4 = Bdr' + Adx'^4. \qquad (57)$$

These, together with (14-17) imply that the Minkowski line element

$$ds^2 = (dx^4)^2 - (dr^2 + r^2(\sin^2(\theta)d\theta^2 + d\Phi^2)), \qquad (58)$$

becomes

$$ds'(\gamma)^2 = G(\gamma)(dx'^4)^2 - (dr'^2 + r'^2(\sin^2(\theta)d\theta^2 + d\Phi^2)), \qquad (59)$$

with

$$G(\gamma) = \{(1-\gamma(r'-x'^4))(1+\gamma(r'+x'^4))\}^2 \qquad (60)$$

$$= \{(1+\gamma(r-x^4))(1-\gamma(r+x^4))\}^{-2} \qquad (61)$$

The metric tensor, $\Gamma(\gamma)$, is diagonal with elements

$$g_{jj}(\gamma) = -G(\gamma), j=1,2,3; g_{44}(\gamma) = G(\gamma). \qquad (62)$$

Minkowski's metric tensor is $\Gamma(0)$.

The $\Gamma(\gamma)$ metric is invariant under the transformations of the rotation group SO(3). If $\gamma$ is not zero, $\Gamma(\gamma)$ and $ds'(\gamma)^2$ are not invariant under the translations and Lorentz transformations of the Poincare group. Thus the general $\Gamma(\gamma)$ metric and line element express the properties of a space-time in which each source of an electromagnetic field reduces Poincare invariance to a rotational invariance about itself.

## 5. Interpretation of Observed Spectral Shifts

The $C_4^{[1]}$ transformations relating primed and unprimed variables define mappings that relate two sets of coordinates, $(r, t, v_r)$ and $(r', t', v'_{r'})$, of the same point. Motions of physical points in Minkowski space will be described by changes in the value of the $(r, t, v'_{r'})$ coordinates of the points, and motions in $G(\gamma)$ space-times will be described by changes in the value of the $(r', t', v'_{r'})$ coordinates of the same physical points.

A point $\Pi$ fixed on an electro-magnetic wave can be assigned the pair of coordinates $r_\pi, x_\pi^4$ and $r'_\pi, x'^4_\pi$. Let $\Pi_1$ and $\Pi_2$ be two such points separated by one wavelength and centered at $\Pi_0$ on a locally monochromatic EM wave propagating in otherwise field-free space-time. Let an observing apparatus at a point **P** momentarily on the wave's source cone have coordinates $r_{ob}, x_{ob}^4, v_{rob}$, and $r'_{ob}, x'^4_{ob}, v'_{rob}$. Let $v$ and $v'$ be frequencies, and $\lambda$ and $\lambda'$ wavelengths measured at **P** in the two systems. Then measuring frequencies in the two sets of coordinates determines the times $\Delta t = \Delta x^4/c = \Delta r/c$, and $\Delta t' = \Delta x'^4/c = \Delta r'/c$, that it takes one wavelength to pass the point **P**. The observations thereby relate time intervals at **P** in the two systems to frequency and wavelength measurements in the two systems via the equations

$$v/v' = \lambda'/\lambda = \Delta x'^4/\Delta x^4. \tag{63,64}$$

We shall assume that the group parameter $\gamma$ is small enough to ensure that approximating the last ratio by $dx'^4/dx^4$ leads to no detectable error. In this approximation

$$\lambda'/\lambda = dx'^4/dx^4 \tag{65}$$

$$= \partial x'^4/\partial x^4 + (\partial x'^4/\partial r) dr/dx^4. \tag{66}$$

Here $dr/dx^4 = v_r$, represents the radial velocity of the observing apparatus at $r$, $t$, in Minkowski space.

FIGURE 2

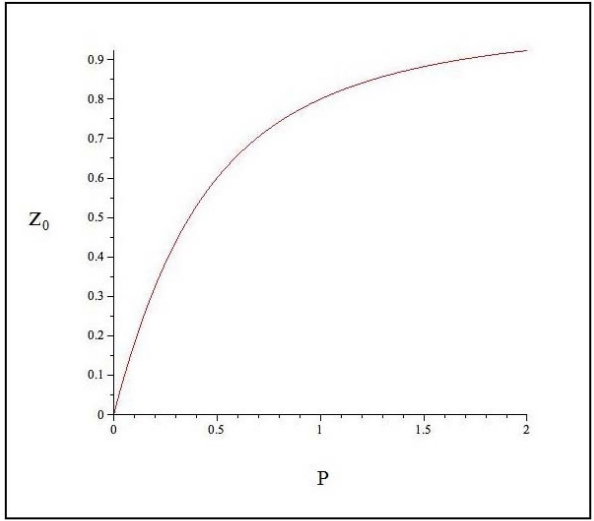

Locally observed velocity-independent redshift of light propagating in G($\gamma$) spacetime. $\rho = \gamma r'$; $r'$ = distance from source. $Z_0 = (\lambda'_{obs}/\lambda_{ref} - 1)$.

Equations (64-66), and (24-30) together imply that

$$\lambda'/\lambda = a' + b' v_r, \tag{67}$$

with

$$a' = (1+\gamma x'^4)^2 + (\gamma r')^2 \tag{68}$$

$$b' = 2(\gamma r')(1+\gamma x'^4). \tag{69}$$

To simplify notation, define

$$\tau = \gamma x^4 = \gamma ct, \ \tau' = \gamma x'^4, \tag{70-72}$$

$$\rho = \gamma r, \ \rho' = \gamma r'. \tag{73,74}$$

On source cones, $\tau = \rho$ and $\tau' = \rho'$.

When expressed in terms of the velocity and position coordinates of an observer's apparatus in G($\gamma$) spacetimes, (67) becomes

$$\lambda'/\lambda = a' + b'(a'v'_{r'} - b')/(a' - b'v'_{r'}). \tag{75}$$

These equations state that $\lambda'/\lambda$ is a function of the time $x'^4$ as well as $r'$. However on a source cone $x'^4 = r'$, so for observations of any particular source

$$b' \to bb = 2(\rho')(1+\rho'), \tag{76}$$

$$a' \to 1+bb. \tag{77}$$

We suppose that $\gamma$ is so small that reference wavelengths, $\lambda_{ref}$ produced and observed by local sources, have the same observable value in Minkowski space as in the G($\gamma$) space-times. Equations (72-77) then relate observables in a G($\gamma$) space-time. We now consider two cases:

1) If $\gamma > 0$, and $v'_{r'} = 0$, then (72-75) predict red shifts that depend solely on $\rho'$. Figure 2 depicts the dependence of $Z_0$ upon $\rho'$. These shifts are given by

$$Z'_0 \equiv (\lambda'/\lambda)\big|_{v'=0} - 1 = \lambda'_{v'=0}/\lambda_{ref} - 1, \tag{78}$$

$$= 2(\rho')(1+\rho')/(1+2(\rho')(1+\rho')), \tag{79}$$

$$\to 2(\rho')(1+\rho') + O((\rho')^4) \tag{80}$$

$$= 2(\rho') + O((\rho')^2). \tag{81}$$

These relations may be compared with the Hubble relation

$$Z = \lambda'_{obs}/\lambda_{ref} - 1 = H_0 R/c + V/c. \tag{82,83}$$

Here $V/c = v$ represents a "peculiar velocity". If it is zero, setting $r' = R$ and $\gamma$ equal to $H_0/2c$ in (83) yields

$$Z = \lambda'_{obs}/\lambda_{ref} - 1 = H_0 R/c = 2\gamma r'. \tag{84-86}$$

This puts (83) in 1:1 correspondence with the Hubble relation.

2) If $\gamma > 0$, and $v'_{r'}$ is nonzero, (76-77) define frequency shifts that are a function of the radial velocity $v'_{r'}$ *and* the position $r'$, of the observer relative to the source. See Figure 3. These shifts are defined by the equations

$$Z' \equiv \frac{Z'_0(1+v'_{r'})}{(1-v'_{r'}Z'_0)} \tag{87}$$

$$= \frac{2\rho'(\rho'+1)(1+v'_{r'})}{\{(1+2\rho'(\rho'+1))(1-v'_{r'})\}} \tag{88}$$

$$= \frac{2\gamma r'(\gamma r'+1)(1+v'_{r'})}{\{(1+2\gamma r(\gamma r'+1))(1-v'_{r'})\}} \tag{89}$$

FIGURE 3

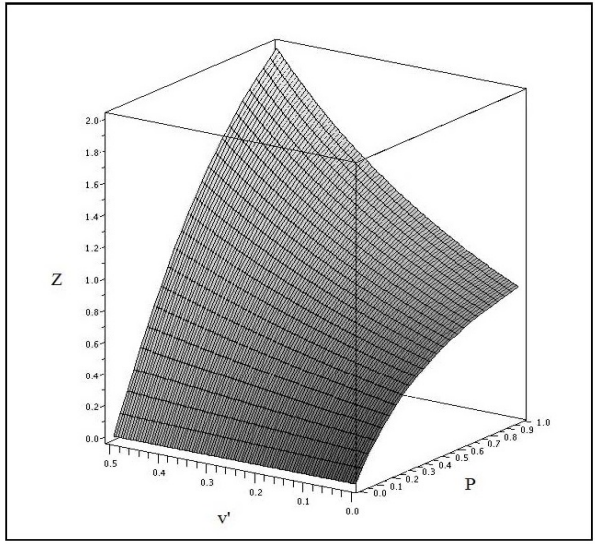

Dependance of total redshift on **P** and velocity **v'**.

When $\rho'$ is <0.1, (89) implies that in the $X'$ system the position and velocity dependence is well approximated by the relations

$$\lambda'/\lambda_{\text{ref}} - 1 = Z' -> 2\gamma r'(1 + v'_{r'}) \quad (90)$$

which are depicted in Figure 4.

FIGURE 4

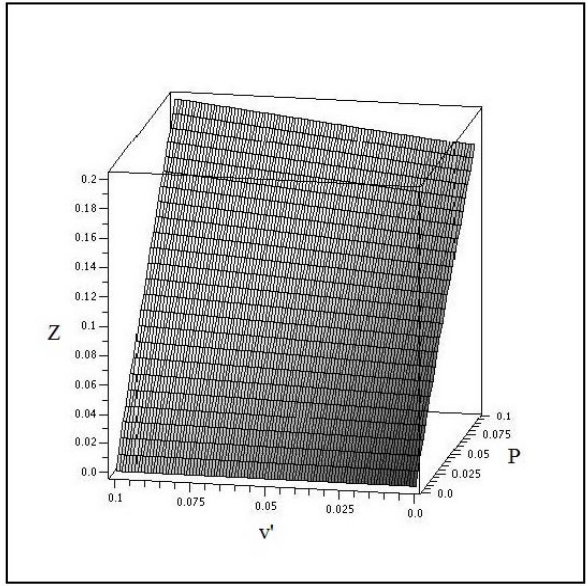

Dependence of redshifts less than 0.2 upon **P** and **v'**.

Equation (90) is to be compared with the result of setting $r' = R$, $2\gamma$ equal to $H_0/c$, and $v'_{r'} = V/c$ in (83) which yields

$$\lambda'/\lambda_{\text{ref}} - 1 = Z_{app} = (H_0/c)r + v_r. \quad (91\text{-}92)$$

*The functional form of this connection between radial velocities and frequency shifts is quite different than that of the connection in* (90). *In the next Section we investigate a relation of* (90) *and* (91-92) *to observations.*

## 6. Observations of Distant Keplerian Motions

In 1914, V. Slipher discovered the existence of internal rotations in nebulae [14]. When F. Zwicky [15] pointed out in 1933 that motions in the Coma cluster of nebulae failed to obey the virial theorem, he suggested that the failure might be due to unobserved "dunkel Materie", or to a failure of the known laws of physics. By 1980, Vera. C. Rubin and her coworkers W. Kent Ford, Jr. and N. Thonnard had established that the motions of stars associated with many spiral galaxies demonstrate the same effect [16].

The failure is now known to be very general, and it is commonly attributed to the presence of unobserved mass and energy, and less commonly, to a modification of Newtonian dynamics. The following paragraphs show that the apparent failure to obey the virial theorem can arise from interpreting observations in a $G(\gamma)$ space-time as observations in Minkowski space.

We follow Peebles [17] and exploit the fact that for circular orbits the mean values in the virial theorem become single values. For a body of mass m and orbital speed $|v|$ orbiting a center with effective mass $M \gg m$, the speed is related to $r$, the radius of the orbit by the well-known equation

$$|v| = (MG/r)^{1/2}. \quad (93)$$

The failure of nebulae to obey (93) is known to appear when observations of luminosities and 21 cm microwave intensities establish that the luminous mass of a nebula lies within a sphere of radius $r_0 < r$. An especially clear example of the phenomenon occurs when the rotational velocities are observed "edge- on". See Figure 5. In most cases, the Doppler shifts then observed are those of velocities indistinguishable from the tangential velocities of bodies moving on circular orbits about the center of mass of the nebulae. These tangential velocities of bodies in Kepler orbits then become radial velocities along the line of sight of the observer.

When the velocities of such motions of massive bodies, and hydrogen, orbiting spiral galaxies have been deduced from observed Doppler shifts, the masses are found to be moving on arcs with radii, $r$, less than the value expected from the mass of the nebulae deduced from luminosities. The velocities also undergo little change as *r* increases. The observed Doppler shifts are thus those that would be obtained if there is sufficient dunkel Materie within the region to keep M*G*/*r* nearly constant as *r* varies over a wide range.

FIGURE 5

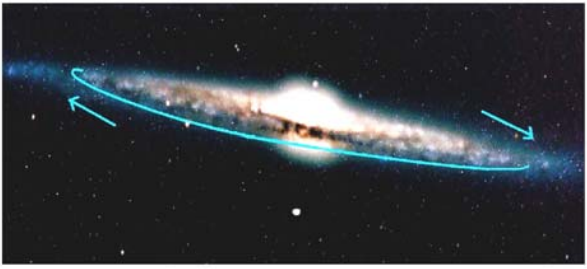

Keplerian trajectory of a body orbiting center of mass of a spiral galaxy.

Equations (90) and (91) suggest that when observations of the motion of bodies in galactic orbit are expressed in $r', x'^4, v'_{r'}$ coordinate systems a very different physical interpretation of the observed motions is implied. This may be shown by examining what would be observed in a G(γ) space-time if a distant orbiting body of mass << than M is moving on a circular Kepler orbit centered on the position of the effective mass M, which is at rest with respect to the observer.

To investigate this question, we consider the orbiting bodies to be origins of source cones, on which the observed orbital tangential velocities become radial velocities of the observer relative to the source. If the observations are being made in an $r', x'^4, v'_{r'}$, system, and the observer is at a distance $r' = D_S$ from the source, then the spectral shifts Z′ arising from $v'_{r'}$ will satisfy the previously noted relation,

$$Z' = 2\rho' \frac{(\rho'+1)(1+v'_{r'})}{\{1+2\rho'(\rho'+1)(1-v'_{r'})\}}, \quad (94)$$

with $\rho' = \gamma r'$, and r' equal to the distance, $D_S$, from the body of mass m to the observer in $G(\gamma)$ space-time. The velocity $v'_{r'}$ is a radial velocity of the observer with respect to the body. Now let $r, r'$ represent radial coordinates of the central mass in the source cone system, and let $v, v'$ represent the value of its tangential velocity in the source cone system. Then, $v' = -v'_{r'}$ and in (94), $v'_{r'}$ and $\rho = \gamma r'$ are related by the equation

$$v'_{r'} = \pm(M'G/r')^{1/2}; \quad (95)$$

in which

$$r' = \exp(\gamma C_4^{[1]})r, \quad (96)$$

$$v'_{r'} = \exp(\gamma C_4^{[1]})v_r. \quad (97)$$

It is revealing to assign values to MG that are those of two of the galaxies investigated in the paper "Dark Matter in Spiral Galaxies" by van Albada and Sancisi [18]. The authors analyze orbital motions observed in five galaxies whose distance from earth ranges from 3.2 to 9.2 mpc. The observed trajectories of the bodies have radii ranging from a few, to 36 kpc, and deduced orbital velocities, $v = V/c$, less than $10^{-3}$. The orbital radii are such a small fraction of the distance $D_S$, that the distance from the observers to the centers of the orbits, $D_G$, is at most 1% greater than $D_S$. As a consequence, the first factor, 2ρ′, in (94), gives Z′ a linear dependence upon the distance $D_G$ from earth to the galaxy, a dependence like that of the Hubble relation (87) if one supposes that 2cγ and $H_0$ have the same order of magnitude.

In the cases being considered, the peculiar velocities have the same magnitude as the velocities of the center of mass of the nebulae relative to the orbiting source. Consequently, for $|v'|$ values of the order of $10^{-4}$ to $10^{-3}$, the factors $(1 \pm v'_{r'})$ in (92) differ from unity by less than 0.1%. Thus, for the reason noted at the end of Section 5, the relation of Z′ to these peculiar velocities is entirely different than that expected from Hubble's relations.

The resulting effect on values of Z′ is displayed in **Table 1**. It contains values of Z′ that would result from measuring spectra of radiation emitted from bodies with velocities v′ orbiting centers with masses equal to those of NGC 2403 and NGC 2841, the galaxies with the smallest, and the largest, mass of the galaxies considered by van Albada and Sancisi. The radii *r'* of the orbits, and the orbital velocities, v′, of the orbiting bodies, are labeled $r_{Newt}$, and $v_{Kep}$ in the table. In it, the hundred-fold change in the radii $r_{Newt}$ changes the local orbital velocities $v_{Kep}$ by a factor of 10. If the red and blue shifts in the G(γ) space-time are interpreted as arising from peculiar velocities in Minkowski space, they produce the values denoted V"Mink". For NGC 2403, over the whole range of $r_{Newt}$, the value of Z′ and V"Mink" vary by 0.02%. For NGC 2841 they change by 0.23%. These results are a direct consequence of the weak dependence of Z′ upon velocity that was noted in Section 5. When Z′ becomes as large as 0.2, increasing the velocity from 0% to 10% of the velocity of light makes a 10% change in Z′.

The results in **Table 1**, taken together with the analysis

**Table 1. Velocities of bodies orbiting the center of mass of spherically distributed masses of galactic magnitude.**

| NGC Object | $D_G$ (Mpc) | MG ($10^{11}$ km) | $r_{Newt}$ ($10^{17}$ km) | $v_{Kep}$ ($10^{-4}$ c) | $v_{Mink}$ ($10^{-4}$ c) | $Z'$ ($10^{-4}$ c) |
|---|---|---|---|---|---|---|
| NGC 2403 | 3.2 | 1.176 | 1.543 | 8.730 | 0.741 h | 0.7412 h |
|  |  |  | 3.086 | 6.173 | 0.741 h | 0.7411 h |
|  |  |  | 6.172 | 4.365 | 0.741 h | 0.7409 h |
|  |  |  | 154.3 | 0.873 | 0.741 h | 0.7407 h |
|  |  |  | Infinite | 0.000 | 0.741 h | 0.7406 h |
| NGC 2841 | 9.0 | 9.555 | 1.543 | 24.89 | 2.088 h | 2.0881 h |
|  |  |  | 3.086 | 17.60 | 2.086 h | 2.0855 h |
|  |  |  | 154.3 | 2.489 | 2.083 h | 2.0834 h |
|  |  |  | Infinite | 0.000 | 2.083 h | 2.0828 h |

The first set of entries apply to a system whose $D_G$ and MG values are those of NGC2403. The second set apply to a system with the $D_G$ and MG values of NGC2841. $v_{Kep}$ is the computed classical velocity of a star in a circular orbit of radius $r_{Newt}$, and center at the center of mass of the indicated galactic system. $Z'$ is the spectral shift that would be observed at distance $D_G$ from the star, measured in G($\gamma$) space-time with $2\gamma = h\, H_0/c$. If the values of Z' are interpreted as arising from peculiar velocities in Minkowski space, they satisfy the relation V = cZ when V << c. These velocities are denoted $v_{Mink}$. The assumed value of $H_0/c$ is $7.5 \times 10^{-25}$ km$^{-1}$, and $0.7 < h < 1$.

in the previous Section, make it evident that if earth-based observations are actually being made in a G($\gamma$) space-time with $2c\gamma$ having a value roughly equal to $H_0$, it is not necessary to modify Newtonian dynamics, or introduce dark matter, to explain the observations: they are a consequence of an unexpected weak dependence of Z' on $v_{Kep}$. This conclusion applies to observed frequency shifts in the radiation received from bodies orbiting a large category of distant galaxies.

The $v_{Kep} \to 0$ entries in **Table 1** illustrate another revealing property of such G($\gamma$) metrics: observations in their space-times do not require the presence of dark energy, or any other type of mass, to explain the spectra of radiation received from objects orbiting at very small velocities in regions arbitrarily distant from centers of gravitational attraction.

## 7. Conclusions

The mere fact that Maxwell's equations allow a $\Gamma(\gamma)$ metric implies that Minkowski's metric, $\Gamma(0)$, cannot be assumed to be the metric of physical gravitation-free space-time.

The analyses in the previous pages establish that observed shifts in the wavelengths of stellar radiations may well imply a relation of frequency shifts to recessionary velocities and peculiar velocities fundamentally different from those codified in Hubble's relations as currently understood. The analyses also establish that the need to introduce dark matter and dark energy, or a modified Newtonian physics, could be eliminated if observed frequency shifts are related to velocities in a space-time in which the $\Gamma(\gamma)$ metric is one with $2c\gamma$ approximately equal to $H_0$.

In the past it has not seemed possible to determine whether Minkowki's metric is valid over distances relevant to Hubble's relation: it is not at all possible to transport clocks and meter sticks to astronomical distances and use them to test Einstein's arguments that lead to Minkowski's metric. However, the value of the group parameter $\gamma$ could be determined by measuring successive return times, as well as frequency shifts within radar wave-pulses sent to receding spacecraft. The uncertainty in the resulting value of the group parameter can be expected to be less than the uncertainty in the value of Hubble's constant [7,8].

If the value turns out to be indistinguishable from zero, the known range of validity of Minkowski's metric will be greatly extended, and much existing astrophysics and cosmology will receive further validation. If the experiments confirm a non-zero value for $\gamma$, they will establish that sources of electromagnetic radiation, in fact, break Poincare and Lorentz invariance in a consequential manner that is both subtle and fundamental.

Such results would have another consequence. Tomilchik [5] and Wulfman [7,8] have both argued that the magnitudes of the "Pioneer Anomalies" roughly correspond to the value of the Hubble constant because the anomalies are primarily due to the assumption that the space-time is Minkowskian. However, Turyshev, *et al.* [19] argue that the anomalies can be explained by a recoil effect which had not been properly accounted for. The proposed experiments would determine the extent to which each explanation is correct.

More importantly, the proposed experiments would effectively establish the metric governing the transmission of electromagnetic waves over distances approaching 100AU.

If it turns out that the value of $2c\gamma$ and Hubble's constant are of the same order of magnitude it will be necessary to revise currently standard interpretations of the spectra of light that has propagated over astronomical distances. The analyses set forth in Sections 5 and 6 establish that the revisions could make it possible to reduce the number of hypotheses currently required to relate astrophysical observations.

Finally, it should be noted that the relations in (29-37) imply that two points moving with the same velocity in the Minkowski coordinate system X have relative accelerations in the X′ system if they are at different distances from the source. This property is reflected in the existence of non-zero forces with components $\partial g_{k,k}/\partial x^j$. Hill's approximation does not make it evident that these accelerations are not constant. The penetrating paper "Conformal Invariance in Physics" by Fulton, Rohrlich, and Witten investigates accelerations produced by conformal transformations [20]. It deals with the mathematics and the physical applications of both conformal function groups and conformal Lie groups. In its final Section the authors state their conclusion, "··· conformal transformations are a special way of describing certain phenomena which in general relativity are accounted for by a restricted class of coordinate transformations". Thus the relation between the $\Gamma(\gamma)$ metric and the metrics of general relativity can be expected to be different than the relation between Minkowski's metric and the metrics of general relativity.

## 8. Acknowledgements

It is a pleasure to acknowledge many clarifying conversations with Andrew S. Wulfman and his assistance with the bibliographic research and preparation of the manuscript.